\documentclass[conference]{IEEEtran}
\IEEEoverridecommandlockouts

\usepackage{cite}
\usepackage{amsmath,amssymb,amsfonts}
\usepackage{algorithmic}
\usepackage{graphicx}
\usepackage{textcomp}
\usepackage{xcolor}

\def\BibTeX{{\rm B\kern-.05em{\sc i\kern-.025em b}\kern-.08em
    T\kern-.1667em\lower.7ex\hbox{E}\kern-.125emX}}

\begin{document}

\title{%
\noindent\fbox{\parbox[c]{0.95\hsize}{\centering\small
\textbf{Notice:} This paper has been published in \textit{2025 IEEE 8th International Conference on Multimedia Information Processing and Retrieval (MIPR)}, San Jose, CA, USA, 06--08 August 2025. DOI: \texttt{10.1109/MIPR67560.2025.00037}
}}\vspace{0.3em}\\
Structural MRI Synthesis for Alzheimer's Disease via Conditional Diffusion on Anatomical Masks}

\author{
    \IEEEauthorblockN{
    Muge Zhang\IEEEauthorrefmark{1},
    Muhammad Ali Khaliq\IEEEauthorrefmark{2},
    Jamal Alsakran\IEEEauthorrefmark{1},
    Byeong Kil Lee\IEEEauthorrefmark{2},
    Jeeho Ryoo\IEEEauthorrefmark{1}
    }
    \IEEEauthorblockA{\IEEEauthorrefmark{1}
    Fairleigh Dickinson University, Vancouver, BC, Canada\\
    m.zhang1@student.fdu.edu, j.alsakran@fdu.edu, j.ryoo@fdu.edu}
    \IEEEauthorblockA{\IEEEauthorrefmark{2}
    University of Colorado at Colorado Springs, Colorado Springs, CO, USA\\
    mkhaliq@uccs.edu, blee@uccs.edu}
}

\maketitle


\begin{abstract}
Recent advances in generative machine learning models have significantly improved medical imaging, offering promising solutions for data augmentation, privacy preservation, and improved model generalization. However, synthesizing high-quality structural MRI data for Alzheimer’s Disease (AD) remains challenging due to the subtle, region-specific, and progressive anatomical changes associated with neurodegeneration. In this paper, we extend the Med-DDPM conditional diffusion model -- originally designed for brain tumor synthesis -- to generate 3D structural MRIs specifically tailored to AD. We adopted Med-DDPM due to its established stability and structural fidelity compared to other generative models, which makes it particularly suitable for capturing the subtle anatomical changes characteristic of AD. Our approach conditions the diffusion process on anatomical segmentation masks derived from the ADNI dataset, incorporating key AD-relevant brain structures into the generation process. We systematically evaluate the quality and utility of the synthetic images by training segmentation models on real, synthetic, and hybrid (mixed) datasets. Experimental results demonstrate that segmentation models trained exclusively on synthetic data achieve comparable Dice scores (0.6532) to those trained on real data (0.6513), while exhibiting significantly enhanced recall. Notably, models trained on hybrid datasets (mixing real and synthetic images) outperform both real and synthetic-only baselines, achieving a Dice score of 0.7244. These findings underscore the successful use of conditional diffusion models for generating anatomically accurate, AD-specific synthetic MRIs, and highlight their potential for enhancing training data availability, improving diagnostic accuracy, and promoting research reproducibility in neuroimaging studies.
\end{abstract}


\section{Introduction}
\label{sec:introduction}
Recent Machine Learning (ML) techniques have led to more accurate diagnosis, segmentation, and treatment planning in various medical domains\cite{Kumar2025}. As the maturity and popularity of ML in medical imaging have grown, obtaining diverse and high-quality training data has become increasingly important. However, in practice, medical imaging datasets -- especially 3D datasets such as Magnetic Resonance Imaging (MRI) -- are limited by privacy regulations, high costs of acquisition, and availability of labeled data\cite{Kragsterman2024}. These challenges are especially evident in neuroimaging, where detailed structural annotations are rare and expensive to obtain, and data sharing is difficult for the aforementioned reasons. 

Synthetic medical image generation has emerged as a strategy to address these limitations. These images are artificially generated based on real images (not obtained from real patients), so free from various restrictions such as privacy. Synthetic imaging methods enable the expansion of training sets and support data augmentation by producing realistic, labeled data, without raising privacy concerns\cite{Ibrahim2024}. Previous works have demonstrated the use of generative models in domains such as brain tumor, multiple sclerosis, and stroke, but applications in neurodegenerative diseases like Alzheimer's Disease (AD) are still not fully explored\cite{Ali2024}\cite{chen2025gpt4o}. In particular, synthesizing realistic AD-affected brain MRIs is challenging because AD pathology is diffuse and progressive, involving subtle atrophy across multiple regions rather than large focal lesions.

In this paper, we explore the use of synthetic 3D brain MRI generation to support diagnostic and segmentation tasks relevant to AD\cite{Dhinagar2024}. Our method adapts the state-of-the-art structural image synthetic method, Med-DDPM, which is a conditional denoising diffusion probabilistic model originally developed for tumor imaging. We extend Med-DDPM for the AD context, focusing on generating synthetic T1-weighted brain MRIs conditioned on brain structure masks from the ADNI dataset\cite{Dorjsembe2024}. In our implementation, we segmented anatomical brain regions that are affected in AD and used these segmentation maps as conditional inputs to train the Med-DDPM model\cite{guo2024synthetic}. The resulting generative model produces full 3D brain MRIs with variable levels of atrophy and other disease-related features characteristic of AD. We then evaluate the quality of the generated synthetic images using a separate segmentation model trained on real data versus synthetic data versus a combination of both\cite{Ali2024}. Our findings show that segmentation models trained on our synthetic images achieve performance comparable to or even better than those trained on real data.

In summary, we make the following contributions:
\begin{itemize}
    \item We extended the Med-DDPM conditional diffusion model to generate 3D brain MRIs specific to AD, conditioning the generation process on segmentation masks of AD-relevant brain structures.
    \item We demonstrate the quality of the resulting synthetic images by evaluating their impact on downstream segmentation model performance, showing that models trained on purely synthetic data can achieve similar performance to models trained on real data.
    \item We illustrate how synthetic MRI generation can serve as a tool for training data augmentation and privacy-conscious data sharing in neuroimaging research, particularly for diseases like AD where data are limited.
\end{itemize}

\noindent The remainder of the paper is organized as follows: Section~ \ref{sec:background} provides background on synthetic image generation and AD imaging; Section~\ref{sec:design} details the architecture and design of our adapted synthesis pipeline; Section~\ref{sec:experimental} describes our experimental methodology; Section~\ref{sec:evaluation} presents experimental results and analysis; Section~\ref{sec:related} discusses recent work in the synthetic medical imaging domain; and Section~\ref{sec:conclusion} concludes the paper with a discussion of future directions.

\section{Background}
\label{sec:background}
Synthetic image generation is the process of creating artificially generated images using generative ML models without using a physical camera or sensor. These images are not captured from the real world, and sometimes do not look anything like the real image to human eyes, yet are very effective in training AI models and simulations. The recent emergence of generative models has facilitated the successful application of synthetic images in various domains such as medical images \cite{Sagers2023}, autonomous driving \cite{Jian2023}, and creative AI \cite{Ramesh2022}. In medical imaging, synthetic data generation has been used for tasks such as data augmentation, lesion and disease simulation, anonymization, image denoising, and even simulating brain morphology changes due to aging\cite{guo2024synthetic}\cite{Billot2023}.



AD poses unique challenges in neuroimaging due to its spatially diffuse pathology. Unlike tumors, which appear as discrete lesions, AD related changes such as cortical thinning and hippocampal atrophy manifest gradually across multiple regions\cite{Mulumba2025}. This lack of a well-localized abnormality complicates both diagnosis and the training of ML models since the disease effects are subtle and distributed. Therefore, it is especially important and valuable to have region-level control in generative models for AD, allowing specific structures to be manipulated or emphasized. However, acquiring large, annotated datasets with detailed anatomical labels for AD is difficult due to privacy constraints, high costs, and the scarcity of expert annotations\cite{Nadeem2024}.

A variety of generative modeling approaches have been explored to overcome these limitations. Prominent categories include Variational Autoencoders (VAEs), Generative Adversarial Networks (GANs), and diffusion models (including latent diffusion)\cite{Vivekananthan2024}\cite{Ye2023}\cite{Friedrich2024}. These models learn the data distribution from a set of neuroimaging scans and can then synthesize new, realistic brain images. For example, VAEs and GANs have been applied to brain MRI synthesis and augmentation tasks\cite{Dhawan2024}. While these approaches have shown promise, each has notable drawbacks. VAEs can suffer from limited image fidelity and issues like posterior collapse, which lead to blurry or less varied outputs\cite{rais2024vae}. GANs, while capable of producing high-resolution images, often face training instability and mode collapse, resulting in difficulty capturing the full diversity of brain appearances. Additionally, GAN training can be sensitive and hard to reproduce, which is problematic for reliable medical image generation\cite{yi2019generative}.


Diffusion-based generative models have recently emerged as a promising alternative that addresses some of these issues\cite{arnold2023stable}. Diffusion models generate images by iteratively denoising random noise, which generally provides greater stability in training and the ability to capture complex data distributions. In the context of medical imaging, conditional and latent diffusion models have achieved state-of-the-art fidelity in synthesis tasks. However, straightforward applications of diffusion models in neurodegeneration research have been limited. For example, latent diffusion models trained on binary AD diagnostic labels can mimic general disease-related features but do not offer fine-grained anatomical control necessary for accurate modeling of AD pathology\cite{hein2024physics}\cite{Dorjsembe2024}. A heuristic approach such as SynthSeg can simulate images from given label maps (segmentations), but it does so without a learned generative model, meaning it may not capture realistic intensity distributions as well as a trained model \cite{Billot2023}.

In comparison to the aforementioned methods, Med-DDPM enables high-fidelity 3D MRI synthesis conditioned on structural masks\cite{Dorjsembe2024}. Med-DDPM is a conditional denoising diffusion probabilistic model that demonstrated effective synthesis of brain tumor MRIs when provided with segmentation masks indicating tumor location\cite{Kim2022}. It achieved competitive segmentation performance using only synthetic training data and showed advantages over GAN-based methods in terms of training stability and output realism.
However, prior implementations of Med-DDPM employed relatively simple conditioning schemes (e.g., binary tumor-versus-background masks), which are less applicable to diseases like AD where structural changes are more subtle and anatomically widespread.
To our knowledge, no prior work has applied structural mask-conditioned diffusion to neuro-degenerative imaging. 

In this study, we extend Med-DDPM to generate AD-specific brain MRIs by conditioning on detailed anatomical segmentation masks relevant to AD pathology. We use segmentation maps produced by FastSurfer\cite{Henschel2020fastsurfer}, a fast and fully-automated deep learning tool for brain MRI segmentation, to provide the structural conditioning. These FastSurfer-generated maps include brain regions known to be affected in AD (such as the hippocampus and ventricles) and are derived from T1-weighted MRIs in the ADNI dataset\cite{Dorjsembe2024}. Without modifying the core Med-DDPM architecture, we retrain the model to synthesize realistic AD-like MRIs conditioned on these anatomical masks. Our goal is to use the resulting synthetic images for training-time data augmentation, and to evaluate whether models trained on synthetic images can perform comparably to those trained on real data. We emphasize that our focus is on structural MRI, and assume that we have access to an automated segmentation method (FastSurfer) to obtain the conditioning masks\cite{Czolbe2023}. Within these bounds, this work offers a targeted, anatomy-aware adaptation of conditional diffusion for neurodegenerative disease research.

\begin{figure}
\centering
\includegraphics[width=\linewidth]{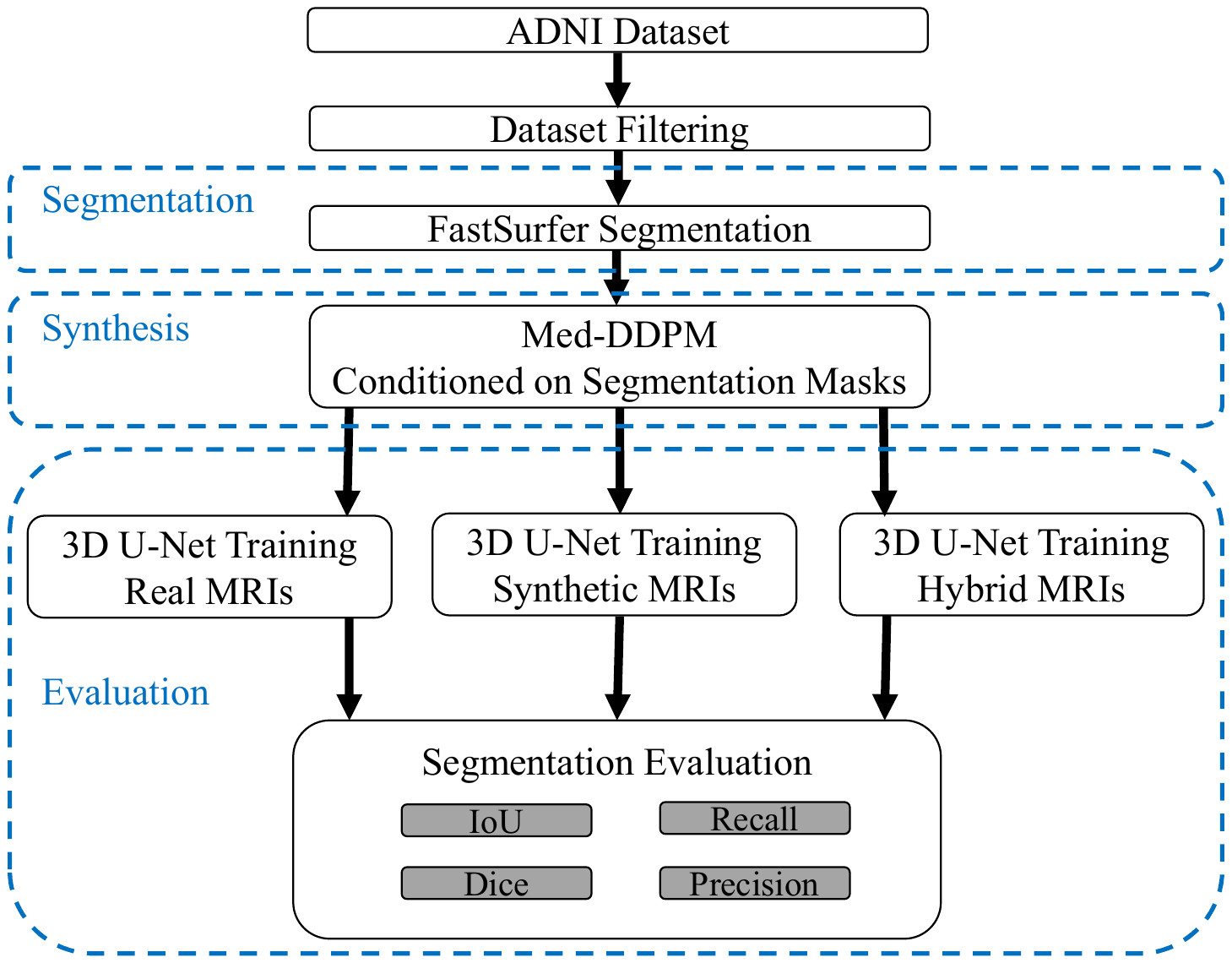}
\caption{The proposed framework for structural MRI synthesis} 
\label{fig1}
\end{figure}

\begin{figure*}[t]
  \centering
  \includegraphics[width=0.8\textwidth]{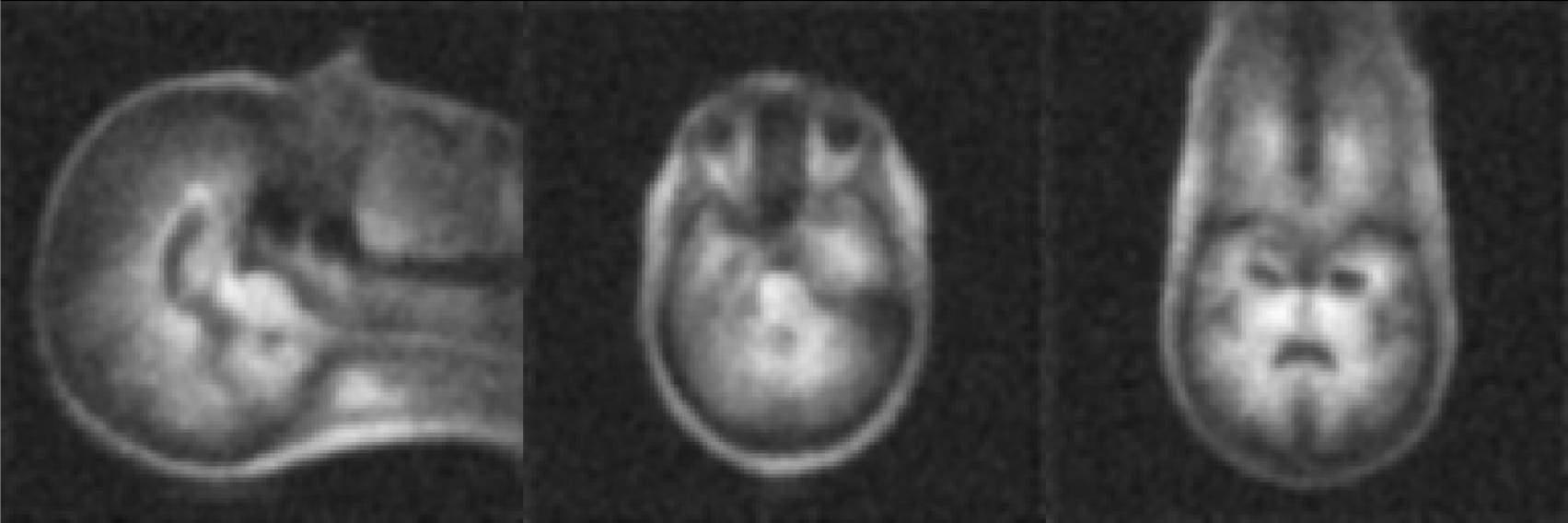}
  \caption{Synthetic T1-weighted brain MRI slices generated by Med-DDPM after 50 training epochs, using our experimental setup.}
  \label{fig:medddpm_synthetic_views}
\end{figure*}

\section{Design}
\label{sec:design}
Our system architecture is designed to determine whether synthetic structural MRI images generated using segmentation-aware diffusion models can serve as a viable alternative to real data for training segmentation models in the context of AD. 
We build our approach on the Med-DDPM framework, adapting it for the AD application. The overall pipeline consists of three major stages (Figure~\ref{fig1}): segmentation, synthesis, and evaluation. 

In the first stage (Segmentation), we segment all T1-weighted MRI images using FastSurfer. It was chosen for its high accuracy, speed, and its use of definitions compatible with the well-established FreeSurfer segmentation protocol. We ran FastSurfer with default parameters in \texttt{--seg\_only} mode (no model fine-tuning). This produces, for each subject, a 3D anatomical label map in the subject’s native space, where each voxel is labeled as one of several brain structures. These segmentation maps serve as structural masks for training the diffusion model in the next stage. In essence, the output of FastSurfer is a mask image where different brain regions (cortex, ventricles, hippocampi, etc.) are encoded by different label values. By using these as inputs, we ensure the generative model is aware of the anatomical layout of each brain.


In the second stage (Synthesis), we use the segmented masks to condition Med-DDPM for image generation. We keep the original Med-DDPM architecture intact but retrain it to generate full-resolution 3D brain MRIs conditioned on the FastSurfer segmentation maps from stage one. During training, the diffusion model learns to map anatomical structures to realistic MRI intensity patterns corresponding to AD-affected brains. This means it captures the structural variability associated with different stages of dementia. For example, AD brains typically show region-specific atrophy, especially in the hippocampus, and enlargement of structures like the lateral ventricles due to tissue loss. The diffusion model learns to translate a given segmentation mask into an MRI volume that reflects such changes (e.g., thinner cortex or enlarged ventricles) in a realistic way. Importantly, although the model is trained on real ADNI scans, it can generate new combinations of anatomical features that were not seen together in any single real subject, while still maintaining plausible anatomy. These novel variations can serve as additional training examples for a segmentation model, potentially helping it generalize better. Figure~\ref{fig:medddpm_synthetic_views} shows representative slices from synthetic 3D T1-weighted MRIs generated by our trained Med-DDPM after 50 training epochs. The synthetic images resemble realistic brain MRIs and exhibit variable degrees of atrophy and ventricle expansion, demonstrating that the model has learned to embed AD-specific traits into the generated images.


In the final stage (Evaluation), we assess the usefulness of the synthetic data by designing three segmentation experiments. We use a standard 3D U-Net architecture~\cite{ronneberger2015unet} for segmenting a particular target structure (or set of structures) relevant to AD (See Figure~\ref{fig1}). The 3D U-Net is widely used in medical image segmentation because it combines low-level and high-level features to accurately delineate anatomical regions, even when those structures are small or have indistinct boundaries. We train three separate U-Net models under different data regimes: one model is trained using only real ADNI MRI images (with their FastSurfer segmentation masks as ground truth labels), another model is trained using only synthetic MRI images generated by our diffusion model (with the corresponding diffusion input masks as labels), and a third model is trained on a hybrid dataset consisting of a mix of real and synthetic images. In all cases, the network architecture and training hyperparameters are kept consistent, so the only difference is the makeup of the training data. This controlled experimental design allows us to isolate the impact of using generated images on segmentation performance. By explicitly conditioning on region-level segmentation during synthesis and using the same segmentation model architecture and evaluation criteria, we conduct a controlled study of the trade-offs between data realism, data diversity, and downstream segmentation performance.



\section{Experimental Setup}
\label{sec:experimental}

All experiments were conducted on a Google Cloud Virtual Machine (a \texttt{a2-highgpu-1g} instance) equipped with 12 vCPUs, 85 GB of memory, and a single NVIDIA A100 GPU (40 GB). This computational setup allowed us to efficiently process high-resolution 3D volumes and to train both the diffusion model and the segmentation models end-to-end within a reasonable time frame\cite{google2023cloud}.

We used T1-weighted structural MRI scans from a curated subset of the ADNI dataset. Subjects were selected to ensure a balanced distribution across diagnosis groups, but genetic or societal factors were not specifically considered. We stratified subjects across three diagnostic categories: cognitively normal (CN), mild cognitive impairment (MCI), and Alzheimer’s disease (AD). We further enforced that each subject contributes only a single scan to the dataset to avoid data leakage from multiple time points.\footnote{In ADNI, many subjects have longitudinal scans taken at multiple visits\cite{jack2008adni}. For our purposes, we randomly picked one scan per subject to ensure that each data point represents a unique individual and to prevent the segmentation model from seeing nearly identical anatomies during training.} After filtering and balancing, we obtained approximately 150 subjects in total to form our training set. We aimed to maintain a roughly equal number of subjects per category to prevent class imbalance, resulting in $\sim$50 subjects per group. This relatively modest dataset size was chosen to keep the training of the high-resolution 3D diffusion model tractable given our computational resources and time constraints. Note that these same subjects (each with a T1 MRI) were used in both the synthesis stage (as input to FastSurfer and the diffusion model) and in the training of the real-data segmentation model (the hybrid model also saw these real scans alongside synthetic ones).



For synthetic image generation, we employed the official Med-DDPM implementation without architectural modifications in order to preserve the model’s proven ability to generate high-quality, anatomy-conditioned 3D medical images. This decision ensures consistent performance with prior work, simplifies reproducibility, and allows us to isolate the effect of AD-specific conditioning without changing model designs. Input images and masks generated by FastSurfer were cropped to 128×128×128 voxels and normalized\cite{isensee2021nnunet}. This ensures compatibility with the Med-DDPM architecture, which requires input dimensions divisible by powers of two for multi-scale processing, while also capturing sufficient anatomical detail for AD-relevant regions without excessive memory overhead. We used four conditioning channels corresponding to AD-relevant anatomical regions including the hippocampus, lateral ventricles, temporal lobe, and cortical gray matter—regions strongly associated with AD progression due to their early involvement in atrophy, fluid accumulation, and structural decline\cite{scherer2016experimental}. These areas are consistently highlighted in clinical and neuroimaging studies as key biomarkers for AD diagnosis and monitoring. No explicit disease severity labels were used. Instead, we rely on the anatomical segmentation masks themselves to reflect pathological changes. For example, hippocampal atrophy or ventricle enlargement is implicitly encoded in the shape and size of the corresponding mask regions. The diffusion model learns to associate these structural patterns with realistic intensity distributions during training. Thus, disease-specific features are captured via the spatial structure of the conditioning masks rather than discrete stage indicators. We trained the model with 155 million trainable parameters for 250 epochs. Each training epoch took approximately 18 minutes on a single NVIDIA A100 GPU, resulting in a total training time of around 75 hours for 250 epochs\cite{dice1945measures}.

To assess the utility of synthetic images, we trained segmentation models using MONAI’s 3D U-Net architecture with approximately 21.3 million parameters. The network consists of four downsampling blocks and uses a hybrid Dice + Cross-Entropy loss\cite{ye2025gpt4o}. We trained three segmentation variants: (1) using real T1w MRIs and masks, (2) using only Med-DDPM-generated synthetic MRIs and their corresponding masks, and (3) using a hybrid dataset (50\% real, 50\% synthetic). Each model was trained for 50 epochs. 



\section{Evaluation}
\label{sec:evaluation}
To evaluate segmentation performance, we compute the Dice similarity coefficient (DSC) for AD-relevant anatomical structures, including the hippocampus, lateral ventricles, temporal lobes, and whole brain volume\cite{jaccard1912distribution}. In addition to Dice, we also used IoU (Intersection over Union), recall, and precision as these metrics offer complementary views of model accuracy. The Dice score measures the overlap between predicted and ground truth segmentations, capturing both false positives and false negatives and offering a balanced assessment of segmentation accuracy\cite{guo2024synthetic}. Therefore, it is a well-used metric for medical image segmentation. IoU quantifies the ratio of the intersection to the union of predicted and actual regions, providing a stricter penalty for misalignment. Precision reflects the proportion of predicted positive pixels that are truly positive, indicating how well the model avoids over-segmentation. Recall measures the proportion of actual positive pixels correctly identified, highlighting the model’s ability to capture relevant anatomical structures without omission\cite{rombach2022high}. 

Performance evaluation of segmentation models is detailed in Table~\ref{tab:segmentation_results}. The evaluation of segmentation results shows that hybrid model performs significantly better than the real and synthetic model, while synthetic model performs slightly better than the real model. The real-only model achieved a validation Dice of 0.6513, IoU of 0.6989, precision of 0.6246, and recall of 0.6739 after 50 epochs. In comparison, the synthetic-only model achieved a Dice of 0.6532, effectively matching the real model. However, metric breakdown reveals contrasting segmentation behavior: the synthetic model yielded much higher recall (0.9743) but lower precision (0.5101), suggesting that it tends to oversegment. In a clinical setting, this behavior can be advantageous for ensuring that no AD-relevant regions are missed, which is critical for early diagnosis. However, this may also introduce false positives, potentially leading to overestimation of pathology and unnecessary follow-up procedures. This likely stems from the structured, mask-conditioned training dataset of Med-DDPM, which encourages completeness and spatial consistency in synthetic MRIs. This could potentially lead to cost of specificity, which means the model is more likely to include areas that do not actually belong to the target structure. While this helps ensure that important regions are not missed, it could also make the results less clear or include extra tissue that is not relevant. The contrast suggests that synthetic data appears to amplify signal around structure boundaries, increasing sensitivity to anatomical regions at risk in AD, such as the hippocampus and lateral ventricles. In practical clinical settings where false negatives are more costly than false positives, this property may be desirable.

\begin{table}[t]
\centering
\caption{Performance summary of segmentation models (real, synthetic, hybrid)}
\label{tab:segmentation_results}
\renewcommand{\arraystretch}{1.2}
\begin{tabular}{|c|c|c|c|c|}
\hline
\textbf{Method} & \textbf{Dice} & \textbf{IoU} & \textbf{Precision} & \textbf{Recall} \\
\hline
Real           & 0.6513 & 0.6989 & 0.6246 & 0.6739 \\
Synthetic      & 0.6532 & 0.4905 & 0.5101 & 0.9743 \\
Hybrid  & \textbf{0.7244} & \textbf{0.5810} & \textbf{0.6333} & \textbf{0.9289} \\
\hline
\end{tabular}
\end{table}

\begin{figure}
    \centering
    \includegraphics[trim = 4cm 17cm 4cm 2.5cm, clip, width=\linewidth]{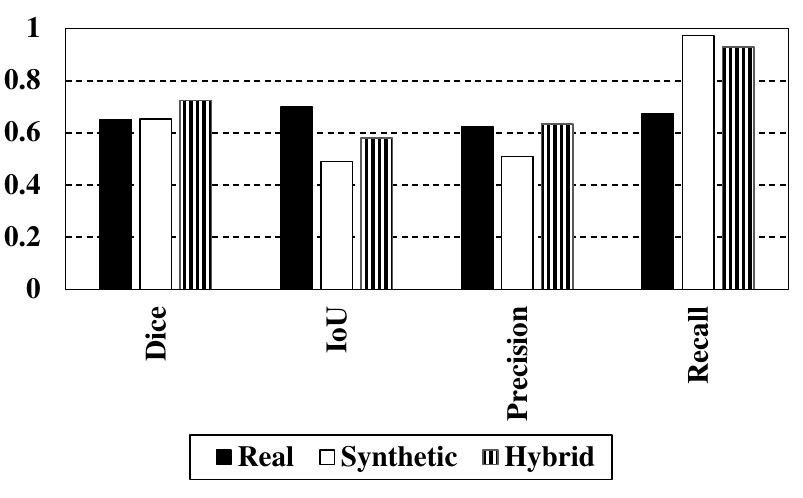}
    \caption{Comparison of segmentation models} 
    \label{fig3}
\end{figure}

The hybrid model substantially outperformed both baselines, achieving a Dice score of 0.7244, IoU of 0.5810, precision of 0.6333, and recall of 0.9289. This demonstrates that combining real and synthetic data yields complementary advantages: the biological realism of real MRIs supports generalization, while the structural diversity and label-aligned consistency of synthetic MRIs reduce variance and promote more robust learning. Visual comparison presented in Figure~\ref{fig3} clearly indicates the hybrid model maintained high recall while improving precision compared to synthetic-only training, suggesting that it preserved the broader structure sensitivity of synthetic images while tempering oversegmentation with real anatomical variability.

\begin{figure}
    \centering
    \includegraphics[trim = 4cm 18.5cm 4cm 2.5cm, clip, width=\linewidth]{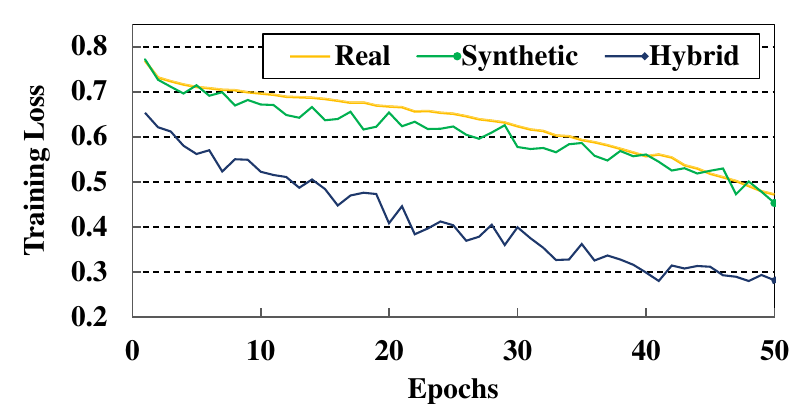}
    \caption{Training loss curves for real, synthetic and hybrid data} 
    \label{fig4}
\end{figure}

Training loss is a measure of how far the model's predictions are from the true labels during training, guiding how the model updates its weights. We used the Dice loss function, which directly optimizes for overlap between predicted and actual segmentation masks. Lower and more stable loss values during training indicate better model convergence, suggesting improved learning quality and greater potential for generalization to unseen MRI scans. Training loss curves reveal clear differences in learning dynamics across the three data regimes as shown in Figure~\ref{fig4}. The hybrid model converges to the lowest final loss and does so more steadily and quickly than the others. This suggests that mixing real and synthetic data yields a more regularized and information-rich training signal, enabling faster generalization and more robust convergence. The synthetic-only model begins with the highest initial loss but quickly catches up to and tracks closely with the real-only model. However, its convergence shows moderate oscillation, particularly in early epochs, reflecting less stability than the hybrid regime. This variation likely stems from differences in the intensity distribution or generative artifacts in the Med-DDPM outputs, despite their strong anatomical alignment. In contrast, the real-only model exhibits more stable but slower convergence, plateauing at a higher loss—possibly due to greater anatomical noise and label variability in real data. These trends align with our segmentation results: synthetic data is sufficient to drive meaningful learning, while hybrid training amplifies the benefits of both sources, producing more stable optimization and superior performance.


Therefore, the success of the synthetic-trained and hybrid-trained models validates the core hypothesis of our work: structurally-conditioned synthetic MRIs generated via diffusion models are not only anatomically plausible but also functionally effective for training segmentation models in neurodegenerative disease contexts. In other words, even though the synthetic images are computer-generated, when they are created with the proper anatomical constraints (segmentation masks), they can provide training signal of comparable quality to actual patient scans. This opens the door for greatly expanding training datasets in a privacy-preserving manner for diseases like AD, which in turn can improve the robustness and generalization of deep learning models for tasks like segmentation and possibly diagnosis.


\section{Related Work}
\label{sec:related}
This section reviews prior work in synthetic medical image generation, focusing on three primary categories relevant to our research: Generative Models in Medical Imaging, Synthetic Data for AD, and Conditional Diffusion Models for MRI.

\noindent\textbf{Generative Models in Medical Imaging}
Early works in generative ML models utilized VAEs to capture latent representations of medical images, offering a foundation for more advanced synthetic image generation techniques \cite{kingma2013auto}\cite{larsen2016autoencoding}. GANs later emerged as powerful tools for synthesizing realistic medical images, particularly useful for data augmentation and anonymization \cite{goodfellow2014generative}\cite{yi2019generative}\cite{shin2018medical}. Recent techniques, including Stable Diffusion and other latent diffusion methods, have further improved image quality, stability, and scalability, enabling diverse applications ranging from lesion simulation to brain morphology modeling\cite{rombach2022high}\cite{Sagers2023}\cite{Ye2023}.

\noindent\textbf{Synthetic Data for AD}
Applying synthetic imaging specifically to AD presents unique challenges due to the disease's diffuse and subtle anatomical changes. Previous research primarily relied on global disease labels to generate synthetic images, limiting the control of anatomical specificity \cite{islam2020gan}. Heuristic methods such as SynthSeg allowed image synthesis from label maps but lacked learned generative capabilities \cite{Billot2023}. Moreover, latent diffusion models trained with binary AD labels have demonstrated the capability to capture general disease-related features but have not offered fine-grained anatomical control necessary for accurate AD modeling \cite{Dhinagar2024}\cite{mulumba2025role}.


\noindent\textbf{Conditional Diffusion Models for MRI}
Conditional diffusion models have recently emerged as promising tools for high-fidelity medical image synthesis. Notably, Med-DDPM demonstrated effectiveness in synthesizing brain tumor MRIs conditioned on structural segmentation masks, providing stability and realism superior to GAN-based methods \cite{Kim2022}. This model has been used primarily for tumor synthesis, demonstrating improved performance for downstream segmentation tasks compared to real datasets alone \cite{Dorjsembe2024}. However, prior implementations of Med-DDPM used simple conditioning schemes (e.g., binary segmentation masks), limiting their application to more complex and anatomically subtle diseases such as AD. Our work extends Med-DDPM by conditioning synthesis specifically on detailed anatomical segmentation masks relevant to AD pathology, enhancing anatomical specificity and improving data utility for segmentation models \cite{Henschel2020fastsurfer}.

\section{Conclusion and Future Work}
\label{sec:conclusion}


In this paper, we presented an extension of the Med-DDPM conditional diffusion model for synthesizing high-quality 3D structural MRI images tailored specifically to AD. By conditioning the generation process on detailed anatomical segmentation masks, our method successfully produces synthetic MRIs that capture subtle, AD-specific anatomical changes. We rigorously evaluated the utility of these synthetic images in a downstream task (brain structure segmentation) and found that models trained on the synthetic data achieved segmentation performance comparable to those trained on real data, and that a hybrid approach combining synthetic with real data yielded even better results. This demonstrates the potential of anatomically-informed diffusion models to alleviate common challenges in medical imaging related to data scarcity, privacy, and annotation costs.

Future work will explore several avenues to extend this research. First, we plan to incorporate longitudinal data and disease progression modeling. By generating sequences of synthetic MRIs over time for the same virtual subject, we could simulate the progression of AD, which may further enhance model training for tasks like change detection or prognostic forecasting. Second, we aim to integrate multimodal imaging data, such as functional MRI or Positron Emission Tomography (PET) scans, into the generative process. Conditioning the diffusion model on multi-modal inputs (or generating multi-modal outputs) could enrich the synthetic dataset, enabling more robust and versatile diagnostic models that leverage complementary imaging information. Third, we will investigate the use of our synthetic data in other downstream tasks, such as classification of AD versus MCI versus CN, to see if the benefits extend beyond segmentation. Fourth, our current approach does not explicitly control or quantify pathological variability across disease stages. Extending the generative model to represent progressive and regionally diverse atrophy patterns will be a key direction for future enhancement. Finally, conducting extensive clinical validation of the synthetic data remains an important step. We plan to incorporate direct image-quality assessments such as Structural Similarity Index Measure (SSIM) or Fréchet Inception Distance (FID) to quantitatively measure anatomical fidelity and intensity realism. We also plan to engage clinicians to assess the realism of the synthetic MRIs and to ensure that any synthetic abnormalities correspond to plausible AD pathology. By verifying that models trained with synthetic data perform well on independent clinical datasets, we can establish greater confidence in the translational value of this approach. Beyond AD, our approach may extend  to other modalities like CT or PET, and even structured generation tasks outside medicine.
In summary, our work demonstrates that conditional diffusion models, when guided by domain-specific anatomical knowledge, can generate realistic and useful medical images for a challenging application. We believe this represents a promising direction for addressing data limitations in medical AI, and we hope it spurs further research into combining generative models with expert knowledge for enhanced data generation in healthcare.

\bibliographystyle{IEEEtranS}
\bibliography{refs}

\begin{thebibliography}{10}
\providecommand{\url}[1]{#1}
\csname url@samestyle\endcsname
\providecommand{\newblock}{\relax}
\providecommand{\bibinfo}[2]{#2}
\providecommand{\BIBentrySTDinterwordspacing}{\spaceskip=0pt\relax}
\providecommand{\BIBentryALTinterwordstretchfactor}{4}
\providecommand{\BIBentryALTinterwordspacing}{\spaceskip=\fontdimen2\font plus
\BIBentryALTinterwordstretchfactor\fontdimen3\font minus
  \fontdimen4\font\relax}
\providecommand{\BIBforeignlanguage}[2]{{%
\expandafter\ifx\csname l@#1\endcsname\relax
\typeout{** WARNING: IEEEtranS.bst: No hyphenation pattern has been}%
\typeout{** loaded for the language `#1'. Using the pattern for}%
\typeout{** the default language instead.}%
\else
\language=\csname l@#1\endcsname
\fi
#2}}
\providecommand{\BIBdecl}{\relax}
\BIBdecl

\bibitem{Ali2024}
M.~H. Ali, M.~Ali, and D.~Koundal, ``{Generative Adversarial Networks (GANs)
  for Medical Image Processing: Recent Advancements},'' \emph{Archives of
  Computational Methods in Engineering}, vol.~32, pp. 1185--1198, October 2024.

\bibitem{arnold2023stable}
\BIBentryALTinterwordspacing
V.~Arnold, ``{Understanding Stable Diffusion: Advantages and Limitations},''
  2023. [Online]. Available:
  \url{https://neuroflash.com/blog/understanding-stable-diffusion-advantages-and-limitations/}
\BIBentrySTDinterwordspacing

\bibitem{Billot2023}
B.~Billot, D.~N. Greve, O.~Puonti, A.~Thielscher, K.~V. Leemput, B.~Fischl,
  A.~V. Dalca, and J.~E. Iglesias, ``{SynthSeg: Segmentation of Brain MRI Scans
  of Any Contrast and Resolution Without Retraining},'' \emph{Medical Image
  Analysis}, vol.~83, p. article 102789, 2023.

\bibitem{chen2025gpt4o}
\BIBentryALTinterwordspacing
S.~Chen, J.~Bai, Z.~Zhao, T.~Ye, Q.~Shi, D.~Zhou, W.~Chai, X.~Lin, J.~Wu,
  C.~Tang, T.~Zhang, H.~Yuan, Y.~Zhou, W.~Chow, L.~Li, L.~Zhu, and L.~Qi, ``{An
  Empirical Study of GPT-4o Image Generation Capabilities},'' \emph{IEEE
  Transactions on Neural Networks and Learning Systems}, 2025. [Online].
  Available: \url{https://ieeexplore.ieee.org/document/2504.05979}
\BIBentrySTDinterwordspacing

\bibitem{Czolbe2023}
S.~Czolbe, A.~V. Dalca, and P.~M. Thompson, ``{Neuralizer: General Neuroimage
  Analysis Without Re-Training},'' in \emph{Proceedings of the IEEE/CVF
  Conference on Computer Vision and Pattern Recognition (CVPR)}, June 2023, pp.
  1234--1245.

\bibitem{Dhawan2024}
K.~Dhawan and S.~S. Nijhawan, ``{Cross-Modality Synthetic Data Augmentation
  Using GANs: Enhancing Brain MRI and Chest X-Ray Classification},''
  \emph{medRxiv}, 2024.

\bibitem{Dhinagar2024}
N.~J. Dhinagar, S.~I. Thomopoulos, and P.~M. Thompson, ``{Generative AI
  Improves MRI-Based Detection of Alzheimer's Disease by Using Latent Diffusion
  Models and Convolutional Neural Networks},'' \emph{Alzheimer's and Dementia},
  vol.~20, no. Suppl. 2, p. article e089958, July 2024.

\bibitem{dice1945measures}
L.~R. Dice, ``{Measures of the Amount of Ecologic Association Between
  Species},'' \emph{Ecology}, vol.~26, no.~3, pp. 297--302, 1945.

\bibitem{Dorjsembe2024}
Z.~Dorjsembe, H.-K. Pao, S.~Odonchimed, and F.~Xiao, ``{Conditional Diffusion
  Models for Semantic 3D Brain MRI Synthesis},'' \emph{IEEE Journal of
  Biomedical and Health Informatics}, vol.~28, no.~7, pp. 4084--4093, July
  2024.

\bibitem{Friedrich2024}
P.~Friedrich, Y.~Frisch, and P.~C. Cattin, ``{Deep Generative Models for 3D
  Medical Image Synthesis},'' in \emph{Generative Machine Learning Models in
  Medical Image Computing}.\hskip 1em plus 0.5em minus 0.4em\relax Springer,
  2024, pp. 255--278.

\bibitem{goodfellow2014generative}
I.~J. Goodfellow, J.~Pouget-Abadie, M.~Mirza, B.~Xu, D.~Warde-Farley, S.~Ozair,
  A.~Courville, and Y.~Bengio, ``{Generative Adversarial Nets},''
  \emph{Advances in Neural Information Processing Systems}, vol.~27, 2014.

\bibitem{guo2024synthetic}
\BIBentryALTinterwordspacing
P.~Guo, D.~Yang, C.~Zhao, and D.~Xu, ``{Addressing Medical Imaging Limitations
  with Synthetic Data Generation},'' \emph{NVIDIA Technical Blog}, 2024.
  [Online]. Available:
  \url{https://developer.nvidia.com/blog/addressing-medical-imaging-limitations-with-synthetic-data-generation/}
\BIBentrySTDinterwordspacing

\bibitem{hein2024physics}
\BIBentryALTinterwordspacing
D.~Hein, A.~Bozorgpour, D.~Merhof, and G.~Wang, ``{Physics-Inspired Generative
  Models in Medical Imaging: A Review},'' \emph{arXiv preprint
  arXiv:2407.10856}, 2024. [Online]. Available:
  \url{https://doi.org/10.48550/arXiv.2407.10856}
\BIBentrySTDinterwordspacing

\bibitem{Henschel2020fastsurfer}
L.~Henschel, S.~Conjeti, S.~Estrada, K.~Diers, B.~Fischl, and M.~Reuter,
  ``{Fastsurfer: A Fast and Accurate Deep Learning Based Neuroimaging
  Pipeline},'' \emph{NeuroImage}, vol. 219, p. 117012, 2020.

\bibitem{Ibrahim2024}
\BIBentryALTinterwordspacing
M.~Ibrahim, Y.~A. Khalil, and M.~Dumontier, ``{Generative AI for Synthetic Data
  Across Multiple Medical Modalities: A Systematic Review of Recent
  Developments and Challenges},'' \emph{IEEE Transactions on Medical Imaging},
  July 2024. [Online]. Available:
  \url{https://ieeexplore.ieee.org/document/2407.00116}
\BIBentrySTDinterwordspacing

\bibitem{isensee2021nnunet}
F.~Isensee, P.~F. Jaeger, S.~A. Kohl, J.~Petersen, and K.~H. Maier-Hein,
  ``{nnU-Net: A Self-Configuring Method for Deep Learning-Based Biomedical
  Image Segmentation},'' \emph{Nature Methods}, vol.~18, no.~2, pp. 203--211,
  2021.

\bibitem{islam2020gan}
\BIBentryALTinterwordspacing
J.~Islam and Y.~Zhang, ``{GAN-Based Synthetic Brain PET Image Generation},''
  \emph{Brain Informatics}, 2020. [Online]. Available:
  \url{https://braininformatics.springeropen.com/articles/10.1186/s40708-020-00104-2}
\BIBentrySTDinterwordspacing

\bibitem{jaccard1912distribution}
P.~Jaccard, ``{The Distribution of the Flora in the Alpine Zone},'' \emph{New
  Phytologist}, vol.~11, no.~2, pp. 37--50, 1912.

\bibitem{jack2008adni}
C.~R. Jack, M.~A. Bernstein, N.~C. Fox, P.~M. Thompson, G.~E. Alexander,
  D.~Harvey, ..., and M.~W. Weiner, ``{The Alzheimer's Disease Neuroimaging
  Initiative (ADNI): MRI Methods},'' \emph{Journal of Magnetic Resonance
  Imaging}, vol.~27, no.~4, pp. 685--691, 2008.

\bibitem{Jian2023}
\BIBentryALTinterwordspacing
Y.~Jian, F.~Yu, S.~Singh, and D.~Stamoulis, ``{Stable Diffusion for Aerial
  Object Detection},'' \emph{IEEE Transactions on Geoscience and Remote
  Sensing}, 2023. [Online]. Available:
  \url{https://ieeexplore.ieee.org/document/2311.12345}
\BIBentrySTDinterwordspacing

\bibitem{Kim2022}
J.~Kim, ``{Med-DDPM: Conditional Denoising Diffusion Model for 3D Brain MRI
  Synthesis},'' \emph{IEEE Transactions on Medical Imaging}, vol.~41, no.~1,
  pp. 123--132, January 2022.

\bibitem{kingma2013auto}
\BIBentryALTinterwordspacing
D.~P. Kingma and M.~Welling, ``{Auto-Encoding Variational Bayes},''
  \emph{Journal of Machine Learning Research}, 2013. [Online]. Available:
  \url{https://www.jmlr.org/papers/v14/kingma13a.html}
\BIBentrySTDinterwordspacing

\bibitem{Kragsterman2024}
P.~Kragsterman, ``{Medical Imaging Research: 2024 Breakthroughs in AI and
  Advanced Technologies},'' \emph{Collective Minds Radiology}, October 2024.

\bibitem{Kumar2025}
R.~R. Kumar, S.~V. Shankar, R.~Jaiswal, M.~Ray, N.~Budhlakoti, and K.~N. Singh,
  ``{Advances in Deep Learning for Medical Image Analysis: A Comprehensive
  Investigation},'' \emph{Journal of Statistical Theory and Practice}, vol.~19,
  p. article number 9, January 2025.

\bibitem{larsen2016autoencoding}
A.~B.~L. Larsen, S.~K. Sonderby, H.~Larochelle, and O.~Winther, ``{Autoencoding
  Beyond Pixels Using a Learned Similarity Metric},'' \emph{International
  Conference on Machine Learning}, pp. 1558--1566, 2016.

\bibitem{Mulumba2025}
J.~Mulumba, R.~Duan, and Y.~Yang, ``{The Role of Neuroimaging in Alzheimer's
  Disease: Implications for the Diagnosis, Monitoring Disease Progression, and
  Treatment},'' \emph{Exploration of Neuroscience}, vol.~4, p. article 100675,
  February 2025.

\bibitem{mulumba2025role}
J.~Mulumba, R.~Duan, B.~Luo, J.~Wu, M.~Sulaiman, F.~Wang, and Y.~Yang, ``{The
  Role of Neuroimaging in Alzheimer’s Disease: Implications for the
  Diagnosis, Monitoring Disease Progression, and Treatment},''
  \emph{Exploration of Neuroscience}, vol.~4, p. 100675, 2025.

\bibitem{Nadeem2024}
S.~Nadeem, P.~Azizan, and I.~Michel, ``{Cortex-Level Brain MRI Generation Using
  Diffusion Models},'' Stanford University, Tech. Rep., 2024.

\bibitem{google2023cloud}
\BIBentryALTinterwordspacing
G.~C. Platform, ``{Google Cloud Virtual Machine Specifications},'' 2023.
  [Online]. Available: \url{https://cloud.google.com/docs}
\BIBentrySTDinterwordspacing

\bibitem{rais2024vae}
\BIBentryALTinterwordspacing
K.~Rais, M.~Amroune, A.~Benmachiche, and M.~Y. Haouam, ``{Exploring Variational
  Autoencoders for Medical Image Generation: A Comprehensive Study},''
  \emph{IEEE Transactions on Medical Imaging}, 2024. [Online]. Available:
  \url{https://ieeexplore.ieee.org/document/2401.12345}
\BIBentrySTDinterwordspacing

\bibitem{Ramesh2022}
\BIBentryALTinterwordspacing
A.~Ramesh, P.~Dhariwal, A.~Nichol, C.~Chu, and M.~Chen, ``{Hierarchical
  Text-Conditional Image Generation with CLIP Latents},'' \emph{IEEE
  Transactions on Pattern Analysis and Machine Intelligence}, vol.~1, no.~2,
  p.~3, 2022. [Online]. Available:
  \url{https://ieeexplore.ieee.org/document/2204.06125}
\BIBentrySTDinterwordspacing

\bibitem{rombach2022high}
\BIBentryALTinterwordspacing
R.~Rombach, ``{High-Resolution Image Synthesis with Latent Diffusion Models},''
  \emph{IEEE Transactions on Pattern Analysis and Machine Intelligence}, 2022.
  [Online]. Available: \url{https://ieeexplore.ieee.org/document/2204.11824}
\BIBentrySTDinterwordspacing

\bibitem{ronneberger2015unet}
O.~Ronneberger, P.~Fischer, and T.~Brox, ``{U-Net: Convolutional Networks for
  Biomedical Image Segmentation},'' in \emph{International Conference on
  Medical Image Computing and Computer-Assisted Intervention}.\hskip 1em plus
  0.5em minus 0.4em\relax Springer, Cham, 2015, pp. 234--241.

\bibitem{Sagers2023}
\BIBentryALTinterwordspacing
L.~W. Sagers, J.~A. Diao, L.~Melas-Kyriazi, M.~Groh, P.~Rajpurkar, A.~S.
  Adamson, V.~Rotemberg, R.~Daneshjou, and A.~K. Manrai, ``{Augmenting Medical
  Image Classifiers with Synthetic Data from Latent Diffusion Models},''
  \emph{IEEE Transactions on Medical Imaging}, 2023. [Online]. Available:
  \url{https://ieeexplore.ieee.org/document/2308.12453}
\BIBentrySTDinterwordspacing

\bibitem{scherer2016experimental}
K.~H. Scherer, ``{Experimental Setup and Methods Development},'' in
  \emph{Grating-Based X-Ray Phase-Contrast Mammography}.\hskip 1em plus 0.5em
  minus 0.4em\relax Springer, 2016, pp. 37--44.

\bibitem{shin2018medical}
H.-C. Shin, ``{Medical Image Synthesis for Data Augmentation and Anonymization
  Using Generative Adversarial Networks},'' in \emph{Simulation and Synthesis
  in Medical Imaging}, 2018, pp. 1--11.

\bibitem{Vivekananthan2024}
\BIBentryALTinterwordspacing
S.~Vivekananthan, ``{Comparative Analysis of Generative Models: Enhancing Image
  Synthesis with VAEs, GANs, and Stable Diffusion},'' \emph{IEEE Transactions
  on Pattern Analysis and Machine Intelligence}, 2024. [Online]. Available:
  \url{https://ieeexplore.ieee.org/document/2408.08751}
\BIBentrySTDinterwordspacing

\bibitem{ye2025gpt4o}
\BIBentryALTinterwordspacing
J.~Ye, J.~Bai, Z.~Zhao, T.~Ye, Q.~Shi, D.~Zhou, W.~Chai, X.~Lin, J.~Wu,
  C.~Tang, T.~Zhang, H.~Yuan, Y.~Zhou, W.~Chow, L.~Li, L.~Zhu, and L.~Qi, ``{An
  Empirical Study of GPT-4o Image Generation Capabilities},'' \emph{IEEE
  Transactions on Neural Networks and Learning Systems}, 2025. [Online].
  Available: \url{https://ieeexplore.ieee.org/document/2504.05979}
\BIBentrySTDinterwordspacing

\bibitem{Ye2023}
J.~Ye, H.~Ni, P.~Jin, S.~X. Huang, and Y.~Xue, ``{Synthetic Augmentation with
  Large-Scale Unconditional Pre-Training},'' in \emph{International Conference
  on Medical Image Computing and Computer-Assisted Intervention}.\hskip 1em
  plus 0.5em minus 0.4em\relax Springer, 2023, pp. 754--764.

\bibitem{yi2019generative}
X.~Yi, E.~Walia, and P.~Babyn, ``{Generative Adversarial Network in Medical
  Imaging: A Review},'' \emph{Medical Image Analysis}, vol.~58, p. 101552,
  2019.

\end{thebibliography}

\end{document}